# A Novel Design for SRAM Bitcell with 3-Complementary-FETs

Xiaoyu Cheng[1,2,3], Yangyang Hu[1,2,3], Tianci Miao[1,2,3], Wenbo Liu[1,2,3], Qihang Zheng[1,2,3], Yisi Liu[1,2,3], Jie Liang[1,2,3], Liang Chen[1,2,3], Aiying Guo[1,2,3], Luqiao Yin[1,2,3], Jianhua Zhang[1,2,3], and Kailin Ren[1,2,3,*]

1 School of Microelectronics, Shanghai University, Shanghai 200444, China
2 Shanghai Collaborative Innovation Center of Intelligent Sensing Chip Technology, Shanghai University, Shanghai 200444, China
3 Shanghai Key Laboratory of Chips and Systems for Intelligent Connected Vehicle, Shanghai University, Shanghai 200444, China
* Correspondence: renkailin@shu.edu.cn

*Abstract*—The complementary field-effect transistors (CFETs), featuring vertically stacked n/p-FETs, enhance integration density and significantly reduce the area of standard cells such as static random-access memory (SRAM). However, the advantage of area scaling by utilizing the CFET structure is hindered by the imbalance in the number of N/P transistors (normally 4N/2P) within the SRAM standard cell. In this article, a novel design that achieves a 6T-SRAM configuration composed of three sets of CFETs is proposed, by stacking two n-FET pass-gate (PG) transistors vertically through a CFET structure. The channel doping concentration, as well as the number of top and bottom nanosheets (NS) are optimized using TCAD simulations, concluding that junctionless accumulation mode (JAM) device instead of inversion mode (IM) device is a better choice for PG and pull-down (PD) transistors in SRAM. An area reduction of 37% in the SRAM standard cell layout is achieved compared with the conventional SRAM layout based on CFET. The predicted performance of the 3-CFET SRAM with n-type channel doping of $1 \times 10^{15}$ cm$^{-3}$ and NS number of '1B4T' demonstrates the greatest overall performances in write margin (349.60 mV) and write delay (54.4 ps). This work provides a promising strategy for the SRAM design in the CFET framework.

*Index Terms*—Complementary Field-effect Transistor (CFET), Static Random Access Memory (SRAM), Junctionless Accumulation Mode (JAM).

## I. INTRODUCTION

A major portion of the total power, performance, and area (PPA) in system-on-chips (SoCs) is dominated by SRAM, a crucial building block in CMOS integrated circuits [1], [2]. Nevertheless, achieving better SRAM integration in SoCs through continuous transistor size reduction suffers difficulties, including severe short channel effects and quantum confinement effects [3], [4]. Despite the excellent electrostatics and gate controllability of Gate-all-around field-effect transistors (GAAFETs), which were proposed after fin field-effect transistors (FinFETs) [5-8], the lateral separation required for n-FET/p-FET in GAAFETs occupies area, hindering further improvements in integration density [9], [10].

By stacking n/p-FETs vertically, the CFET structure addresses this issue, leading to 1.5 to 2-fold gain in integration density beyond traditional GAA technologies [11].

Despite the aforementioned advantages, there are still several issues on attaining the full potential of CFET to further reduce the SRAM standard cell area, such as routing congestion and lithography scheduling constraints. Many methods have been proposed to address these issues. In [12], the routing congestion related to the requirement to cross-couple two inverters in SRAM is resolved by recessing the top device and enabling the write line (WL) gate contact to access the bottom device at the cell border.

While further shrinking the area, the implementation of Buried Power Rail (BPR), the Backside Contact (BSC) Power Delivery Network (PDN), and other structural boosters successfully reduced the congestion in SRAM top signal tracks and enhanced device performances [13–17]. The photolithography constraints in the gate cut (GC) process step are alleviated by using gate dielectric isolation walls rather than GC, which lowers the height of SRAM cells [18].

However, considering that 6T SRAM is not an N/P complementary logic structure due to the existence of two pass-gate (PG) n-FETs, current SRAM construction scheme composed of four CFETs with two p-FETs invalidated fails to exploit the full potential of CFET structures for further scaling and improvement of integration density.

In this article, a novel design for the 6T-SRAM standard cell composed of three sets of CFETs is proposed, in which two PG n-FETs are stacked vertically in a CFET structure, and junctionless accumulation mode (JAM) device instead of inversion mode (IM) device is utilized to PG and pull-down (PD) transistors. This design further decreases the area of the SRAM standard cell without compromising the circuit performance. The layout design, fabrication process, and mixed-mode simulation results are covered in Section II. The compact model and the parasitic parameter extraction is delivered in Section III. In Section IV, the circuit simulation results by HSPICE are investigated, with conclusions reached in Section V.

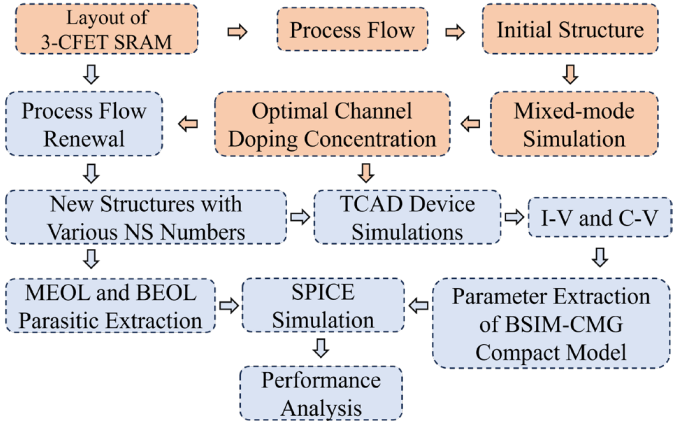

Fig. 1. The evaluation flow in this work.

TABLE I
LAYOUT DESIGN RULES ASSUMED AND OTHER PARAMETERS

| Parameter | Quantity | Value |
|---|---|---|
| $W_{NS}$ | Nanosheet width | 6 nm |
| $T_{NS}$ | Nanosheet thickness | 5 nm |
| $CPP$ | Contact Poly Pitch | 42 nm |
| $FP$ | Fin pitch | 28 nm |
| $L_g$ | Gate length | 12 nm |
| $GC$ | Gate cut | 26 nm |
| $GE$ | Gate extension | 5 nm |
| $W_{BPR}$ | BPR width | 25 nm |
| $N_{VS}$ | Nanosheet Vertical Spacing | 9 nm |
| $L_{sp}$ | Spacer length | 5 nm |
| $T_{ox}$ | Thickness of low-k oxide | 0.4 nm |
| $T_{hk}$ | Thickness of high-k oxide | 1.5 nm |
| | M0 pitch | 16 nm |
| | M1 pitch | 21 nm |

## II. 3-CFET SRAM DESIGN

### A. Evaluation Flow

The evaluation flow in this work is demonstrated in Fig.1. Firstly, the initial structure of the 3-CFET SRAM is constructed based on the proposed layout and process flow using TCAD simulation tool SEMulator3D. This structure serves as the basis for the extraction of the TCAD device structures of pull-up (PU), PD, and PG transistors. Subsequently, mixed-mode simulation is employed to investigate the impact of channel doping concentration on the performance of SRAM and to obtain the optimal channel doping concentration. Following this, by varying the nanosheet (NS) number in the top and bottom device, 3-CFET SRAM structures with interconnects are produced. Based on these structures and the optimized channel doping concentration, TCAD simulations of PU, PD, and PG transistors are conducted to obtain their *I-V* and *C-V* characteristics for parameter extraction based on BSIM-CMG compact models. In the meantime, the middle-end-of-line (MEOL) and back-end-of-line (BEOL) parasitic parameters are extracted. Finally, the BSIM-CMG compact model as well as the parasitic parameters are written into a netlist for SPICE circuit simulations to conduct parasitic-aware device-circuit co-optimization and thoroughly analyze the performance of the 3-CFET SRAM.

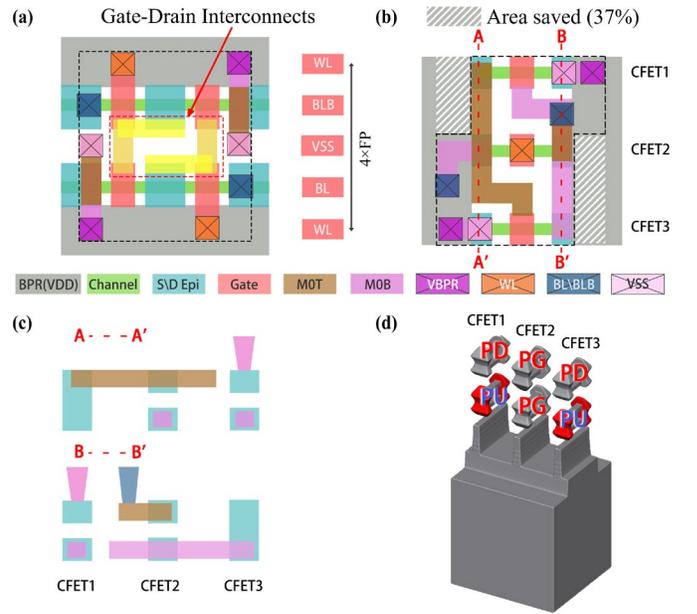

Fig. 2. (a) Layout of a conventional SRAM based on CFET [12]. (b) Layout of 3-CFET SRAM with 37% area saved compared with (a). (c) Detail of connections in A-A' and B-B' cross-sections of the layout of 3-CFET SRAM. (d) Schematic diagram of the positions of PU, PD, and PG transistors in 3-CFET SRAM.

### B. Layout Design

Fig. 2(b) depicts the layout design of a 6T SRAM using three sets of CFETs, with complete design criteria listed in Table 1. Instead of the conventional design with mixed-horizontal-and-vertical distributions depicted in Fig. 2(a) [12], 3-CFET SRAM utilizes a layout with three groups of CFETs placed vertically. Additionally, placing the BPRs used for VDD connectivity perpendicular to the NSs simplifies the VBPR distribution and avoids the additional pitch caused by the restriction of lithography process when BPRs are parallel to the fins [13]. The schematic of the positions of PU, PD, and PG transistors is illustrated in Fig. 2(d). Two inverters in SRAM are formed by the n/p-FETs stacked on top and bottom of the CFET1 and CFET3. Two n-FETs, stacked through the CFET2, serve as two PG transistors. As shown in Fig. 2(c), the common drain contacts of the two sets of inverters are placed diagonally and connected to the sources (or drains) of the two PGs through the M0 bottom (M0B) and top (M0T) metal interconnects. The interconnects between the bottom PG and Bit Line (BL) requires extra space to be extended outward since the M0T takes up the top space of the CFET2. This extended metal line is part of the M0B. The area of the 3-CFET SRAM is indicated by the black dotted line in Fig. 2(b), since the BPR on either side of it is not completely occupied and the vacant space can be shared with the nearby SRAM units on the left and right side. In summary, it is calculated that the layout area of the 3-CFET SRAM is 37% smaller than that of the conventional SRAM based on CFET.

### C. Fabrication Process Simulation

The fabrication process flow of the 3-CFET SRAM in TCAD simulations, which is based on the monolithic CFET integration process [19-21], is depicted in Fig. 3(a). The images after several key process steps are presented in subfigures (i)-(viii). Three key processes are explained as follows. First, after step 8,

The SiGe:B source/drain regions are epitaxially grown on the exposed Si. Then, the top epitaxial SiGe:B source/drain is etched away, and a Si:P is epitaxially re-grown on the top source/drain regions. Since the bottom device of CFET2 is also a JAM n-FET, the entire source/drain epitaxy of CFET2 needs to be etched away before re-growing Si:P. The final result is shown in subfigure (vi) of Fig. 3(a). Second, the gate work function metals of n/p-FETs are deposited using the vertically stacked dual metal gate technique described in [20], as shown in step 14. Finally, after the gate cut process in step 15, the gate metal of CFET2 needs to be completely removed, since p-type work function metal was deposited for the gates of the bottom devices of all CFETs in step 14, and then n-type work function metal is deposited to form the CFET structure with two n-FETs.

It is worth noting that the dopant type should be consistent for distinct devices in the same layer, since in-situ doping during the epitaxial growth is utilized in SiGe/Si stacking in the 3-CFET SRAM fabrication process [22]. As shown in Fig. 3(b) the channel doping of the bottom device in the 3-CFET SRAM is set to n-type doping. As a result, the PG device at the bottom is with n-type channel doping, making it a JAM device. The doping distribution along the A-A' direction is shown in Fig. 3(d). Meanwhile, the PU transistor is set as IM type, and its specific doping distribution is shown in Fig. 3(e). As shown in Fig. 3(b), the channel doping of the top device is also set to n-type doping, which makes the top PG and PD both JAM devices.

### D. Device Mixed-Model Simulation

In this investigation, the following physical models are enabled and taken into consideration during the TCAD simulations. The Inversion and Accumulation Layer Mobility Model, the Doping Dependence Model, and the Thin Layer model are utilized and coupled with the quantum confinement effects, given the 5 nm thickness of the NS. The Hurkx band-band tunneling model, and the Auger and Shockley-Read-Hall recombination models are applied in the recombination models. Stress is applied to the device in the channel direction using the Piezo model, with stress values of 0.5 MPa for n-FET, and 1.5 GPa for p-FET [23]. Through calibration with experimental data [24], the gate metal work function is modified to get symmetric $V_{th}$ with $|V_{th}|$ of 0.3 V, based on the conventional channel doping concentrations of p-type ($1\times10^{16}$ cm$^{-3}$) and n-type ($1\times10^{18}$ cm$^{-3}$) [25].

To validate the accuracy and feasibility of the TCAD simulations, Fig. 4(b) demonstrates the calibration results that are obtained by calibrating the TCAD simulation settings with experimental data [24]. Fig. 4(a) shows the process of TCAD calibration based on published experimental data [24], where the device structure and key characteristics are extracted. Then, the physical model parameters are adjusted to fit the TCAD simulation results with the experimental data.

Next, in order to obtain the optimal channel doping concentration of the top and bottom devices, a mixed-mode TCAD simulation is carried out on the read static noise margin (RSNM) of 3-CFET SRAM, in which the influence of interconnects is excluded and only the individual influences of the properties of the PU, PD, and PG devices are investigated. The reason for choosing RSNM as the key indicator is that the static performance of SRAM is often evaluated by RSNM, and the dynamic characteristics of SRAM are significantly affected by parasitic resistance and capacitance, which are not proper indicators for the optimization of channel doping concentration. Fig. 5 shows the mixed-mode simulation results for two different combinations of NS numbers ("xByT", where x represents the number of NSs in the bottom and y represents the number of NSs in the top). The optimal RSNM, exceeding 119 mV for 2B2T, and exceeding 109 mV for 1B4T, is achieved with n-doping concentrations of $1\times10^{15}$ cm$^{-3}$ for both bottom and top devices.

On the other hand, as shown in Fig. 6(a), the top device which is a JAM type has a larger on-state current ($5.61\times10^{-5}$ A for a 4 NSs device) compared with the IM type device, while the off-state leakage current is almost the same ($1.89\times10^{-10}$ A). Therefore, with this design, a nine-stage ring oscillator with a top JAM-type n-FET on the same wafer with 3-CFET SRAM exhibits a higher frequency (0.443 GHz) in the mixed-mode simulation, as shown in Fig. 6(b). So, one of the most optimal channel concentrations is chosen to n-doping of $1\times10^{15}$ cm$^{-3}$ for both the top and bottom devices.

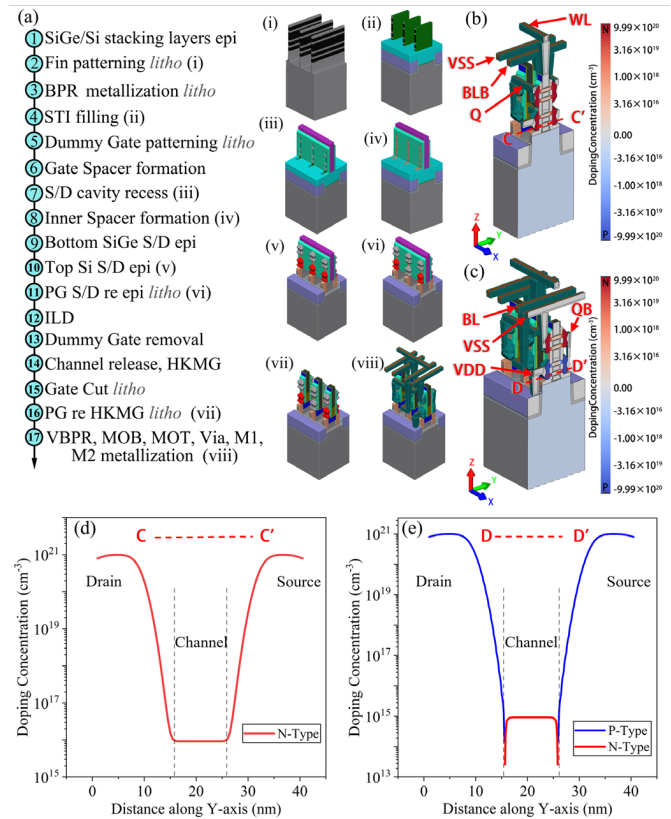

Fig. 3. (a) Schematic diagram of the process simulation flow for fabricating SRAM standard cells under the 3-CFET strategy, and subfigures (i)-(viii) present the results after key steps with the inter-layer dielectric (ILD) hidden. (b) Cross-sectional view and doping distribution along the A-A' direction at the CFET2 channel. (c) Cross-sectional view and doping distribution along the B-B' direction at the CFET3 channel. (d) Detailed doping concentration profile for the C-C' cross-section. (e) Detailed doping concentration profile for the D-D' cross-section.

                                                                                                                                          

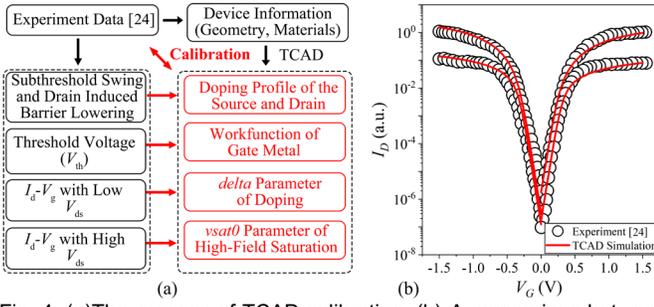

Fig. 4. (a)The process of TCAD calibration. (b) A comparison between TCAD simulation and experimental data [24] for the transfer characteristics of p- and n-FET.

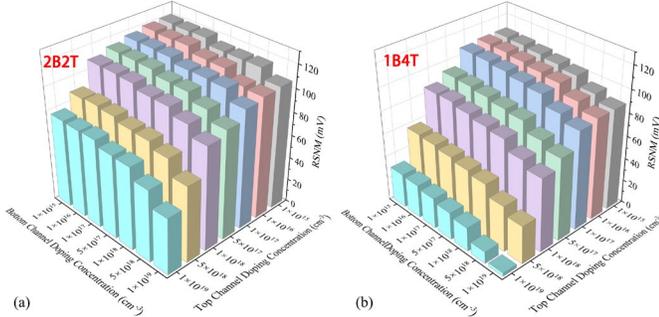

Fig.5. The mixed-mode simulation results on RSNM for 3-CFET SRAM structures with two different combinations of NS numbers: (a) 2B2T, (b) 1B4T.

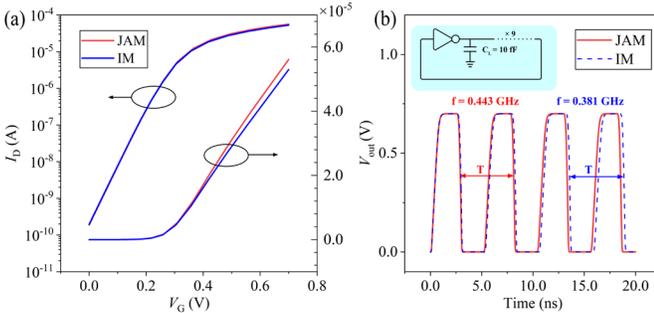

Fig. 6. (a) Transfer characteristic curves of IM and JAM n-FETs with a 4-NS channel. (b) Mixed-mode simulation results of a nine-stage ring oscillator with a top IM n-FET and JAM n-FET.

## III. PARASITIC AND COMPACT MODEL EXTRACTION

### A. Parasitic Capacitance and Resistance Extraction

3-CFET SRAM with 16 various combinations of the top and bottom NS numbers is generated by altering the process, in which the number of Si channel layers in the SiGe/Si stack is increased from one to four. The MEOL and BEOL parasitic capacitances of all structures are extracted using the Raphael tool. Additionally, the MEOL parasitic capacitance between the common gates of the CFETs and their connections is extracted concurrently with the overall parasitic capacitance extraction. The results of the parasitic capacitance extraction for various combinations of bottom and top NS numbers are displayed in Fig. 7(a). The VDD node interconnect is mainly composed of the BPR and VBPR, which are located near the bottom of the overall structure and far from other interconnects. Consequently, its parasitic capacitance value is not significantly affected by an increase in NS number. The WL, Q, and QB interconnects are with large parasitic capacitance because they are positioned in the center of the SRAM structure and connected to the gates of the three CFETs. A significant increase in parasitic capacitance is also brought about by the addition of NS number, with an average increase of 5.48% per NS.

Considering that the parasitic resistance of metal interconnects cannot be directly extracted in SPX simulations [26], the parasitic resistance in the proposed 3-CFET SRAM is estimated by calculating the length of metal interconnects multiplied by the unit resistance of interconnects, where the unit resistance of Ru is $R_{Ru} = 795$ Ω/μm and the unit resistance of Mo for BPR is $R_{BPR} = 50$ Ω/μm [27][28].

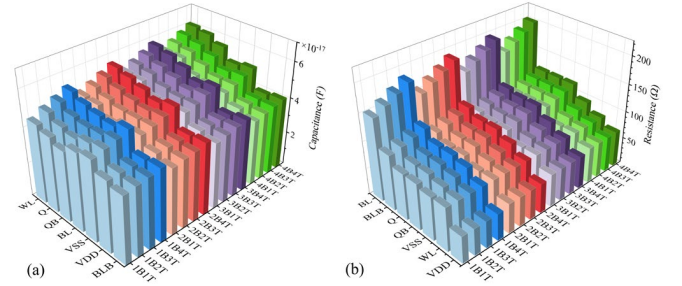

Fig. 7. (a) Results of parasitic capacitance extraction for various combinations of bottom and top NS numbers. (b) The estimation results of parasitic resistance.

Fig. 7(b) shows the extraction results for parasitic resistance in various structures. It is evident that = the resistance of BL is significantly higher than that of other interconnects. This is because BL needs to be connected from M1 to M0B where the bottom PG is located, and in the worst case, the length of the M1 metal line where BL is placed is almost equal to the width (y-axis) of the SRAM. Another notable finding is that the parasitic resistance of the VDD and WL interconnects, which are situated at opposite ends of the structure, remains almost unaffected by the addition of middle NSs. In contrast, the parasitic resistances related to other nodes rise as the NS number increases.

### B. Compact Model Extraction

Based on the C-V and I-V characteristics of the devices in TCAD simulations, the parameters in BSIM-CMG 110.0 models for each individual device are extracted. First, the flag parameter *GEOMOD* in BSIM-CMG 110.0 is set as 2 to characterize the quad-gate of NSs devices, and the NSs structure is equivalently treated as a Fin, with the parameter *TFIN* used to define the width of the NS, and the parameter *HFIN* used to define the total thickness of the NSs. Only models of the top JAM n-FET (PD) and the bottom IM p-FET (PU) are necessary to be extracted, with PG model being the same as PD model, considering that both the top PG and top PD are JAM devices with same structural parameters and that the *I-V* characteristics of bottom and top PG devices of the CFET are symmetric by design. Fig. 8 compares the TCAD simulation and the modeling results of the C-V characteristics and transfer characteristics of n-FET and p-FET devices with 1-4 NS channels to illustrate the quality of the model extraction.

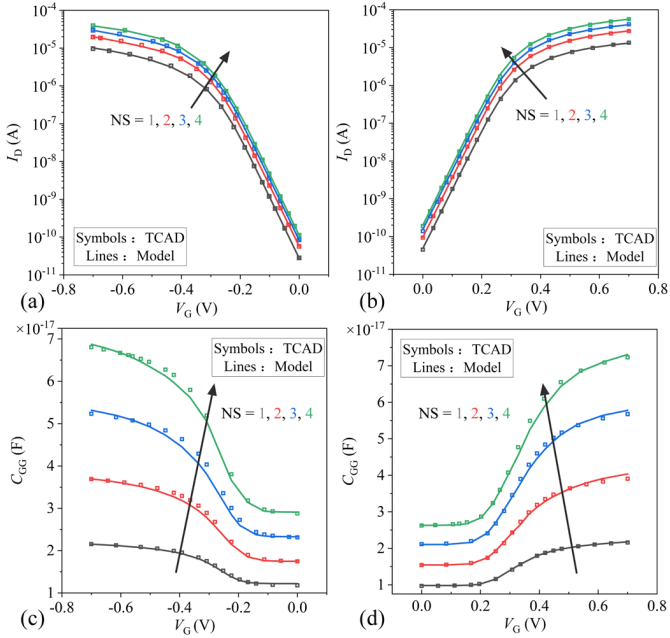

Fig. 8. Comparison between TCAD simulation and modeling results: Transfer characteristics of(a) p-FET and (b) n-FET; Capacitance characteristic of (c) p-FET and (d) n-FET.

## IV. 3-CFET SRAM PERFORMANCE PREDICTION AND ANALYSIS

### A. Read Operation

The RSNM and read delay of the 3-CFET SRAM are extracted to evaluate how NS number influences the read operation. A DC sweep from 0 V to 0.7 V is applied to the nodes Q and QB during the extraction process, while the bias on BL, BLB, and WL remains $V_{dd}$ (0.7 V). Fig. 9(a) illustrates the results of the RSNM extraction. The RSNM is noticeably superior when the top NS number is larger than or equal to that of the bottom. These structures exhibit an average RSNM of 105.78 mV, with the '4B4T' structure achieving the highest RSNM of 118.30 mV. In contrast, the '4B1T' structure performs the poorest among all combinations, allowing only 20.70 mV of RSNM for data reading from SRAM. This result is attributed to the fact, that the driving power of the PG transistor positioned on the bottom is greater than that of the PD when the bottom NS number exceeds the top NS number, as illustrated in Fig. 8(b). Consequently, during a read operation, the Bit line (BL) charges the node QB, which is stored as a '0'. Consequently, the potential of QB is increased, which in turn lowers the RSNM of SRAM.

BL and BLB are floating nodes initially set to 0.7 V, and WL is provided with a pulse signal with a rise time of 2 ns to switch on both PGs while extracting the read delay, which is defined as the difference between the time that WL rises to half of the $V_{dd}$ (0.35 V) and the time that the BL/BLB level reduces by 200 mV. The two PGs cannot always be kept symmetric in the design due to variations in NS number. Therefore, the read delay obtained is the time it takes to read the storage '0' via both the BL and BLB pathways.

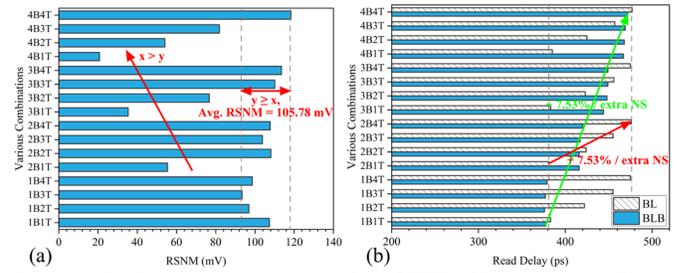

Fig. 9. Performance analysis of 3-CFET SRAM with different combinations of NS numbers in the bottom and top: (a) RSNM. (b) Read delay in BL and BLB paths.

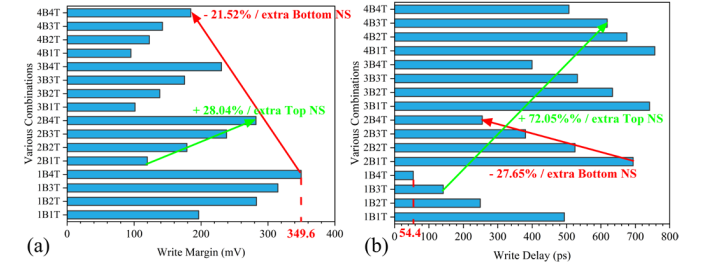

Fig. 10. Performance analysis of 3-CFET SRAM with different combinations of the NS numbers in the bottom and top: (a) Write margin. (b) Write delay.

Fig. 9(b) displays the read delay extraction results in BL and BLB pathways. For every extra top NS, the read delay increases by 7.53% while reading stored data from the BL path linked to the bottom PG. It is noteworthy that increasing the bottom NS number has virtually no impact on the read delay of the BL path when the number of underlying NSs stays constant. This phenomenon is similarly observed when the '0' potential stored at node Q is read by the BLB path. The reason is that the parasitic capacitance related to the PG transistor causes charge to accumulate during the read operations, increasing the time it takes to read data. Moreover, as illustrated in Fig. 8(d) and 7(c), the average increase in the parasitic capacitance of the gate of the PG transistor ($C_{gg}$) with every additional NS (1.69×10⁻¹⁷ F), is significantly greater than that in the parasitic capacitance of the interconnects (1.22×10⁻¹⁸ F). Meanwhile, the read delay of the BL path increases with the increase of NS number when the NS number on the top and bottom is equal, because the parasitic resistance and capacitance of BL are greater than those of BLB. Therefore, by lowering the quantity of NSs on the bottom and top, the read delay in the 3-CFET SRAM can be optimized. The '1B1T' structure, with the fewest NSs, can achieve a minimum read delay of 383 ps.

### B. Write Operation

To evaluate the write performance of the 3-CFET SRAM, the write margin and write delay are investigated. DC sweep is imposed on WL to obtain the write margin [29]. Fig. 10(a) presents the write margin of 3-CFET SRAM with various combinations of the NS numbers in the bottom and top. The best write margin of 349.60 mV is achieved in the '1B4T' structure. It is evident that the write margin decreases when the bottom NS number increases, when the top NS number stays constant. The write margin drops by 21.52% on average for every added bottom NS. The reason is that more bottom NSs improve the PU's driving ability, thus charging the node that stores '1' to a higher potential. Consequently, it becomes more challenging to discharge the node to '0' potential through the PG and PD to accomplish the write operation, which lowers the write margin.

The write margin rises by 28.04% on average for every additional top NS with a constant bottom NS number. The underlying mechanism is that adding more top NSs will boost the driving capability of PDs and cause the node that stores '1' to discharge to a lower potential, since both PD devices are at the bottom. As a result, it will be easier to discharge the node that stores '1' to '0' via any PG. Consequently, the write margin may be increased by adding more top-layer NSs.

A pulse signal with 2 ns rising time is imposed to WL to activate both PGs and execute the write operation to extract the write delay, which is defined as the time interval between WL increases to 1/2 of $V_{dd}$ and the storage '1' and '0' reverses. The influence of various NS number combinations on write delay extraction results are illustrated in Fig. 10(b). An average increase in write delay of 72.05% is observed with each increment in the number of bottom NSs. Conversely, the write delay is decreased by an average of 27.65% for each additional top NS.

## V. Conclusion

This article proposes a novel 3-CFET design for SRAM standard cells, which saves 37% of the area by stacking two n-FET PG transistors within the CFET structure. For this design, a co-optimization of the channel doping concentration and NS number is carried out. It is verified by simulations that the 3-CFET SRAM exhibits an RSNM of greater than 119 mV when the channel is lightly n-doped with $1\times10^{15}$ cm$^{-3}$ for both the bottom and the write channels. As for the influence of NS numbers, the write margin improves by 28.04%, the read delay increases by 7.53%, and the write delay decreases by 27.65%, on average for every added top NS, while the write margin decreases by 21.52% and the write delay rises by 72.05% on average for each extra bottom NS. Therefore, it is predicted that 3-CFET SRAM with the '1B4T' structure achieves the optimal performances when considering both read and write capabilities.

## Acknowledgment

This work was supported by the National Natural Science Foundation of China under Grant 62204150, the Natural Science Foundation of Shanghai under Grant 23ZR1422500, and the Special Funds for Promoting High-quality Industrial Development in Shanghai (JJ-ZDHYLY-01-23-0004).